\documentclass{INTERSPEECH2023}
\usepackage{array,multirow}
\usepackage{boldline}
\usepackage{authblk}

\usepackage{lipsum}

\usepackage[activate]{microtype}

\usepackage[dvipsnames]{xcolor}
\usepackage{soul}

\title{A Preliminary Study on Augmenting Speech Emotion Recognition\\ using a Diffusion Model}
\name{$^*$Ibrahim Malik$^1$,  $^*$Siddique Latif$^2$, Raja Jurdak$^2$, and Bj\"{o}rn W.\ Schuller$^{3,4}$}

\address{
  $^1$Emulation AI\\
  $^2$Trusted Networks Lab, Queensland University of Technology (QUT), Australia
   $^3$GLAM -- Group on Language, Audio, \& Music, Imperial College London, UK\\
  $^4$Chair of Embedded Intelligence for Health Care and Wellbeing, University of Augsburg, Germany}
\email{siddique.latif@qut.edu.au}

\begin{document}
\interspeechcameraready
\maketitle
\def\thefootnote{*}\footnotetext{These authors contributed equally to this work}\def\thefootnote{\arabic{footnote}}
\begin{abstract}
In this paper, we propose to utilise diffusion models for data augmentation in speech emotion recognition (SER). In particular, we present an effective approach to utilise improved denoising diffusion probabilistic models (IDDPM) to generate synthetic emotional data. We condition the IDDPM with the textual embedding from bidirectional encoder representations from transformers (BERT) to generate high-quality synthetic emotional samples in different speakers' voices\footnote{synthetic samples URL: \url{https://emulationai.com/research/diffusion-ser.}}. We implement a series of experiments and show that better quality synthetic data helps improve SER performance. We compare results with generative adversarial networks (GANs) and show that the proposed model generates better-quality synthetic samples that can considerably improve the performance of SER when augmented with synthetic data.

\end{abstract}

\section{Introduction}
Speech emotion recognition (SER) aims at enabling machines to perform emotion detection using deep neural networks (DNNs) models \cite{schuller2018speech,latif2023transformers}. SER has a wide range of applications in customer centres, healthcare, education, media, and forensics, to name a few \cite{latif2020speech}. Various studies have explored different deep learning (DL) models including deep belief networks (DBN) \cite{latif2018transfer}, convolutional neural networks (CNN) \cite{yenigalla2018speech}, and long short-term memory (LSTM) networks \cite{wollmer2012analyzing} to improve the performance of SER systems. However, the SER performance is hindered by the unavailability of
larger labelled datasets \cite{alimulti}. 
Developing high-quality emotional datasets can be a time-consuming and costly process \cite{latif2021survey}.


Data augmentation is considered an effective method to generate synthetic samples to tackle the data scarcity problem in SER. Various studies (e.g., \cite{aldeneh2017using,latif2022multitask,baird2021emotion}) in SER have shown the effectiveness of audio data augmentation techniques including SpecAugment \cite{park2019specaugment}, 
speed perturbation \cite{ko2015audio}, 
and noise addition \cite{lakomkin2018robustness}. However, speech variations like speed perturbations do not change the semantic content and they may have an effect on emotional expressions. Another approach is to utilise generative models including a conditional 
generative adversarial network (GAN)
\cite{mirza2014conditional}, Balancing GAN \cite{mariani2018bagan}, StarGAN \cite{choi2018stargan}, and CycleGAN \cite{zhu2017unpaired} to generate emotional features to augment the SER system (e.\,g.,  \cite{chatziagapi2019data,rizos2020stargan}). Studies have found that synthetic data by generative models can help improve the performance of SER systems. However, vanilla GANs face convergence issues due to smaller emotional corpora and are unable to produce high-quality emotional synthetic features \cite{sahu2018enhancing,latif2020augmenting}. To address these issues, we propose to use diffusion models to generate synthetic data to augment the SER system \cite{shahid2023generative}. In contrast to GANs, diffusion models provide better training stability and produce high-fidelity results for audio and graphics \cite{dhariwal2021diffusion,huang2022fastdiff}. 
To the best of our knowledge, this paper is the first to explore diffusion models for SER.

The key contribution of this paper is the use of improved denoising diffusion probabilistic models (IDDPM) to generate synthetic data to augment the training of the SER system. In order to generate high-quality synthetic samples, we condition the IDDPM with text embedding from bidirectional encoder representations from transformers (BERT) \cite{devlin2019bert}. We present a comprehensive analysis by evaluating the SER system in (i) within the corpus and 
(ii) 
cross-corpus settings on 4 publicly available datasets. We empirically show that synthetic data generated by the proposed framework considerably improves the SER results compared to recent studies. We will publicly release the code of our models. 

\section{Related Work}
Various data augmentation techniques are used to improve SER performance. 
Speed perturbation \cite{ko2015audio} is a popular technique that is widely used in SER to generate augmented data. For example, studies \cite{latif2019direct,aldeneh2017using} use speed perturbation as a data augmentation method and evaluate it in SER. Based on the results, they show that data augmentation helps improve SER performance. SpecAugment \cite{park2019specaugment} is another augmentation technique that was proposed for automatic speech recognition (ASR). Studies \cite{baird2021emotion,latif2022multitask} explore the SpecAugment technique in the SER 
domain 
and show that 
the data SpecAugment 
improves the generalisation and performance of the systems. Recently, the mixup \cite{zhangmixup} data augmentation technique is also being explored in SER. Mixup generates the synthetic sample as a linear combination of the original samples. Latif et al.\ \cite{latif2020deep} use mixup to augment the SER system in order to achieve robustness. Based on the results, they found that augmentation helps improve generalisation by generating diverse training data. Other studies \cite{aldeneh2017using,abdelwahab2018study} also use data augmentation to improve the performance of SER by increasing the training data. However, these studies do not use generative models to generate synthetic data. 

Generative models aim to generate new data points with some variations by learning the true data distribution of the training set. GANs are popular generative models due to their ability to learn and generate data distributions. Different studies have explored GANs 
in 
SER. For instance, Sahu et al.\ \cite{sahu2018enhancing} explored a vanilla GAN and a conditional GAN to generate a synthetic emotional feature vector from a low-dimensional (2-d) feature space. They use a support vector machine (SVM) as a classifier and computed the results on both real and synthetic data. They found that the vanilla GAN faces convergence issues due to the smaller emotional corpus, however, a conditional GAN could generate better synthetic features that help improve the SER performance. Recently, the authors in \cite{latif2020augmenting} attempted to augment their GAN with mixup \cite{zhangmixup} augmentation and achieved better SER performance. In contrast to these studies, we use the diffusion model to generate high-quality synthetic samples to augment the SER system. 

\section{Proposed Approach}
\begin{figure}[!ht]
\centering
\includegraphics[width=0.41\textwidth]{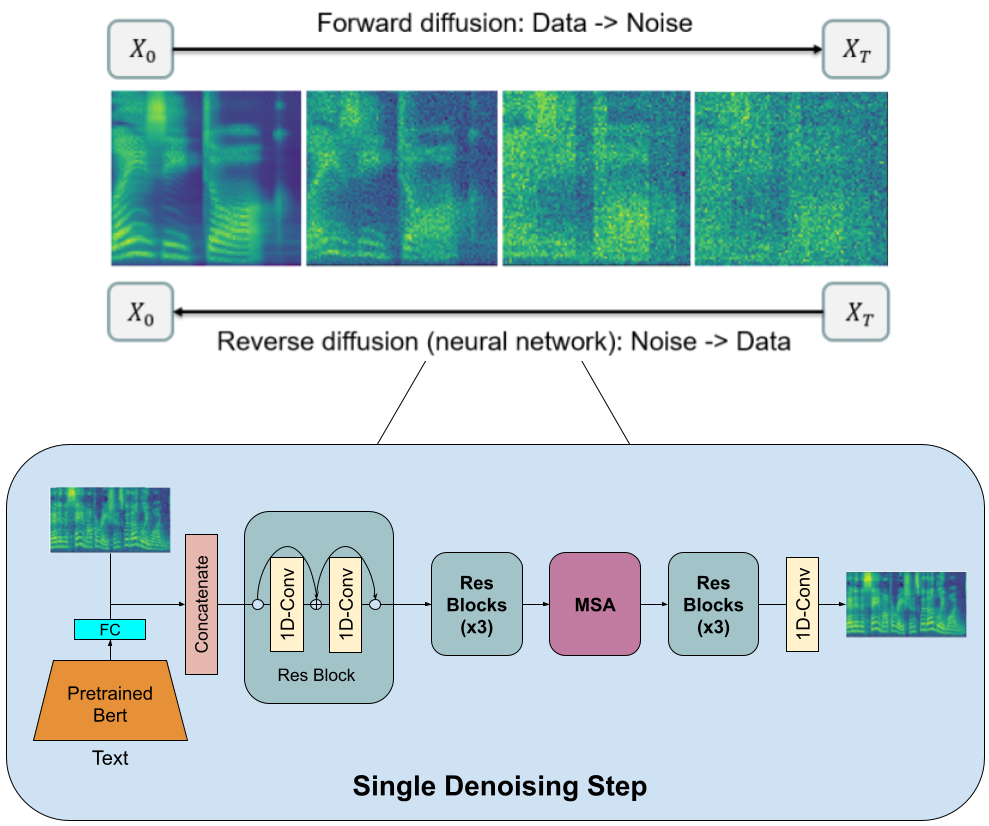}
\vspace{-.6em}
\caption{Illustration of the forward and reverse diffusion process. In the forward phase, we add Gaussian noise on each timestep until the sample becomes an isotropic Gaussian distribution. In the reverse phase, we estimate the noise for each timestep using a neural network and denoise the corrupted sample.}
\vspace{-1.5em}
\label{fig:Model}
\end{figure}

We use improved denoising diffusion probabilistic models (IDDPM) to generate emotional data. Figure \ref{fig:Model} shows the proposed model that takes spectrogram conditioned on text embedding to generate synthetic emotional spectrograms. The details of the proposed framework are presented next. 

\subsection{Diffusion for Emotional Data Synthesis}
Diffusion models are state-of-the-art generative models inspired by non-equilibrium thermodynamics. They are fundamentally different from other popular generative models including GANs and 
variational auto-encoders (VAEs) 
\cite{kingmaauto}. They define the diffusion process as a Markov chain by slowly adding random noise to the input data for a total of $T$ times and learn to reverse this process by reconstructing the desired data samples from the noise. Various diffusion model architectures have been proposed, however, we utilise the improved denoising diffusion probabilistic models (IDDPM) \cite{nichol2021improved}, which is an extended version of denoising diffusion probabilistic models (DDPM) \cite{ho2020denoising}. The main motivation for the utilisation of IDDPM is its improved log-likelihoods and that it requires fewer timesteps to generate high-fidelity outputs. 


Given an emotional data point $x_0$ sampled from a real data distribution, we add Gaussian noise for $T$ timesteps using a forward noising process $q$ that provides latent 
for each timestep. 
The sample $\mathbf{x_T}$ becomes isotropic Gaussian noise for a large $T \to \infty$. This is highlighted in Figure \ref{fig:Model} which shows that an initially clean Mel-spectrogram ($x_0$) is completely transformed into Gaussian noise after adding noise for each timestep $t$. The noise is according to a schedule of $\beta_t$ and the forward noising process can be defined as: 
\begin{align}
q(\mathbf{x}_t \vert \mathbf{x}_0) &= \mathcal{N}(\mathbf{x}_t; \sqrt{\bar{\alpha}_t} \mathbf{x}_0, (1 - \bar{\alpha}_t)\mathbf{I}),
\end{align}
$\text{where } \alpha_t = 1 - \beta_t \text{ and } \bar{\alpha}_t = \prod_{i=1}^t \alpha_i$. We illustrate a single denoising step in Figure \ref{fig:Model} that shows the approximation of noise at each timestep done by a neural network $p_{\theta}$ :
\begin{align}
    p_{\theta}(\mathbf{x_{t-1}}|\mathbf{x_{t}}) := \mathcal{N}(\mathbf{x_{t-1}}; \mu_{\theta}(\mathbf{x_t},t); \Sigma_{\theta}(\mathbf{x_t},t )),
\end{align}
where $\mu_{\theta}(\mathbf{x_t},t)$ and $\Sigma_{\theta}(\mathbf{x_t},t )$ are the mean and variance parameters respectively for the noisy sample at timestep $t$. 

If we know the exact reverse distribution $q(x_{t-1}|x_t)$, we can sample $x_t \sim \mathcal{N}(0,I)$ and generate $x_o$ by running the process in reverse. As explained by Ho et al.\ \cite{ho2020denoising}, 
a denoising neural network can be trained to predict $\mathbf{x_{t-1}}$ from $\mathbf{x_{t}}$ at timestep $t$ using the following:
\begin{align}
\mathbf{x}_{t-1} &= \mathcal{N}(\mathbf{x}_{t-1}; \frac{1}{\sqrt{\alpha_t}} \Big( \mathbf{x}_t - \frac{1 - \alpha_t}{\sqrt{1 - \bar{\alpha}_t}} \boldsymbol{\epsilon}_\theta(\mathbf{x}_t, t) \Big), \boldsymbol{\Sigma}_\theta(\mathbf{x}_t, t)).
\end{align}
Essentially, the network learns to predict the mean and variance parameters of noise for $t-1$ at each $t$. This noise is then subtracted from the $\mathbf{x_{t}}$ and we obtain $\mathbf{x_{t-1}}$. Ho et al.\ \cite{ho2020denoising} kept $\beta_t$ fixed as constants and set $\boldsymbol{\Sigma}_\theta(\mathbf{x}_t, t) = \sigma^2_t \mathbf{I}$, and $\sigma_t$ is either set  to  $\beta_t$ or $\tilde{\beta}_t = \frac{1 - \bar{\alpha}_{t-1}}{1 - \bar{\alpha}_t} \cdot \beta_t$.
We use IDDPM \cite{nichol2021improved} that learns  $\boldsymbol{\Sigma}_\theta(\mathbf{x}_t, t)$ and select a cosine-based noise schedule instead of a fixed $\beta_t$. This helps reduce the time for sampling by utilising a strided sampling schedule. The sampling is updated after every $[T/S]$ step, which reduces the sampling time from $T$ to $S$. The new sampling schedule for generation is $\{\tau_1, \dots, \tau_S\}$, where $\tau_1 < \tau_2 < \dots <\tau_S \in [1, T]$ and $S < T$. We generate the synthetic samples by training the model for both forward and reverse processes. The training process and model configuration are explained next.

\subsection{Model Configuration and Training}


In our IDDPM model, we select a commonly used modified version of the UNet \cite{ronneberger2015u} for the denoising process. In the modified UNet, a self-attention layer between the bottleneck and CNN layers is used. For our experiments, we observed better results in terms of audio synthesis by using 1-dimensional CNN layers instead of the more commonly used 2-dimensional CNN for image generation. This also helps reduce the memory footprint of the model and speed-up training. 
To ensure that each synthetic sample had a consistent speaker emotion and does not contain nonsensical gibberish as speech, we conditioned our denoising network on a representation of target samples. We passed the corresponding text with emotion and speaker information of each target sample to a pre-trained bidirectional encoder representations from transformers (BERT) \cite{devlin2019bert} model to get the representation and extrapolated it simply by using fully connected layers with a linear sigmoid unit (SilU) activation followed by a self-attention layer. This representation was concatenated with the input at each timestep $t$. This modified input is passed through a block of 8 1-dimensional CNN layers with 1536 filters, each. Each layer has a 
SiLU activation and we provide residual connections between consecutive layers which we call res-blocks. A self-attention layer is provided afterwards and we add 3 more res-blocks. A final 1-d CNN layer is added to bring the number of filters back to 80.

For our training procedure, we use a cosine-based noise schedule and choose 4000 diffusion steps to learn the variance along with the mean of noise for each timestep $t$. We trained the model on an Nvidia Rtx 3090 GPU with a batch size of 64 for about 120,000 steps. During inference, we control the emotion and speaker's voice in the output samples using the condition vector.


\section{Experimental Setting}

\subsection{Datasets Details}
We selected four publicly available popular emotional datasets for SER evaluations. The details of these datasets are presented below.

\textbf{IEMOCAP:} The interactive emotional dyadic motion capture (IEMOCAP) \cite{busso2008iemocap} is a multimodal dataset containing English dyadic
conversations. The dataset spans over 10 sessions and two speakers for each session. The annotation is performed by 3-4 assessors in 10 emotions. For consistency with previous studies \cite{bao2019cyclegan,latif2020augmenting,sahu2018enhancing}, we use four emotions (angry, sad, happy, and neutral) for our experiments.  The total samples for these selected emotions are 5531.

\textbf{MSP-IMPROV:} For cross-corpus evaluation, we select the
``acted corpus of dyadic interactions to study emotion perception'' (MSP-IMPROV) \cite{busso2017msp}. Similar to IEMOCAP, this corpus also contains recordings of English dyadic conversations. It consists of six sessions, with utterances from two speakers per session. In total, there are 7,798 utterances with four emotions: neutral, sad, angry, and happy. All samples from the corpus are used for our experiments. 

\textbf{CREMA-D:} The crowd-sourced emotional multimodal actors dataset (CREMA-D) \cite{cao2014crema} is a data set of 7,442 clips from 91 actors. We select this corpus for cross-corpus evaluations across datasets having different distribution and recording scenarios. The clips in this corpus are from 48 male and 43 female actors between the ages of 20 and 74 coming from a variety of races and different ethnicities. 
In our experiments, we only use four emotions for our experiments. 



\textbf{RAVDESS:} The Ryerson audio-visual database of emotional speech and song (RAVDESS) \cite{livingstone2018ryerson} is another popular multimodal database. This corpus is gender balanced consisting of 24 professional actors, vocalising lexically-matched statements in a neutral North American accent. Similar to CREMA-D, we select this data for cross-corpus evaluations. We select four emotions from this data similar to other datasets used in this paper.  


\subsection{Pre-processing and Input Representation}
A popular method in SER is to represent speech as Mel-spectrograms. Likewise, we compute the Mel-spectrograms using a short-time Fourier transform of size 1024, 256 hop-size, and a window size of 1024. We select the frequency range of 0-8\,kHz and extract 80 Mel frequency bands scaled linearly in the range of $[-1,1]$. To cater for the varying audio length, we use a segment-based approach for training the model as used in \cite{yenigalla2018speech}. We use the segment length of three seconds. The larger utterances are segmented and the smaller ones are zero-padded. 
This results in a Mel-spectrogram of shape (80 x 256) for each segment.

\section{Experiments and Evaluations}
In this section, we present the results of our diffusion model and our SER.

\subsection{Synthetic Data}
\begin{figure}[!ht]
\centering
\includegraphics[width=0.45\textwidth]{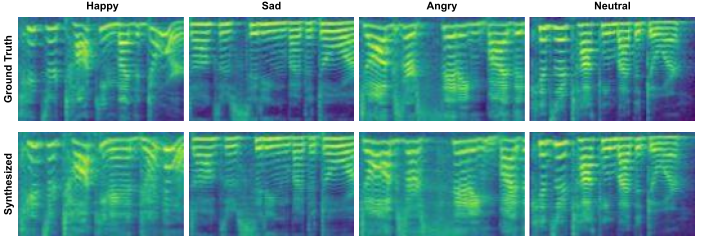}
\caption{Comparing ground truth samples of IEMOCAP data with synthetic samples generated using our proposed model. }

\label{fig:synGt}
\end{figure}

We generate synthetic data using IDDPM for different emotions. In Figure \ref{fig:synGt}, we plot synthetic Mel-spectrograms for each emotion and compare them with the corresponding ground truth samples. To evaluate the quality of our synthetic audios, we use a pre-trained HiFi-GAN \cite{kong2020hifi}, a vocoder that can synthesise high-fidelity audio waves from Mel-spectrograms. We choose a sampling rate of 22.5\,kHz as 
HiFi-GAN 
is trained on this  sampling rate.
We calculate the mean absolute difference (MAD) between the ground truth samples and our synthesised samples for IEMOCAP and the results are presented in Table \ref{table:mad}. This shows that our synthetic data have very small variations from the real data for all emotions. Our synthetic samples for happy and neutral have slightly high MAD, however, we are achiving better score in contrast to the Bao et. al. \cite{bao2019cyclegan}. These variations in synthetic data help the speech emotion classifier learn from diverse information and improve the SER performance that we highlight in the next experiments.   
\begin{table}[!ht]
\centering
\scriptsize
\caption{Mean of absolute difference for different emotions. }
\begin{tabular}{|c|c|}
\hline
\textbf{Emotion}               & \textbf{Mean of Absolute Difference} \\ \hline
Happy                          & 0.0141                                                \\ \hline
Sad                            & 0.0096                                                \\ \hline
Angry                          & 0.0097                                                \\ \hline
Neutral                        & 0.0143                                                \\ \hline
Total                 & \textbf{0.0476}                                                \\ \hline
Bao et. al. \cite{bao2019cyclegan} ($\lambda^{cls} = 2$) & 0.0490                                                \\ \hline
\end{tabular}
\vspace{-1.5em}
\label{table:mad}
\end{table}
\subsection{Speech Emotion Classification}
In this section, we perform SER to empirically evaluate the quality of synthetic data. We augment the training data and present the results for within-corpus and cross-corpus settings. We implement a 
convolutional neural network (CNN)
and bidirectional LSTM (CNN-BLSTM) based classifier for SER. We use three 1-D CNN layers with rectified linear layers as activations to learn high-level representations from input Mel-spectrograms. These representations are then passed to BLSTM layers to learn the emotional context. The final output from the BLSTM layer is fed to a fully connected layer that gives an output vector equal to the number of emotions classes. We use batch normalisation to speed up training and add a dropout of 0.1 between the CNN layers and 0.2 between the LSTM layers. We train each model for 100 epochs with a learning rate of $10^{-5}$ and a batch size of 64. We select all these parameters using the validation set. 
To have a fair comparison with previous studies \cite{latif2020augmenting,bao2019cyclegan,sahu2018enhancing}, we use a leave-one-speaker-out scheme and results are presented 
by the field's standard measure 
unweighted average recall. 

\subsubsection{Within-Corpus}
\begin{figure*}[!ht]
\centering
\includegraphics[width=\textwidth]{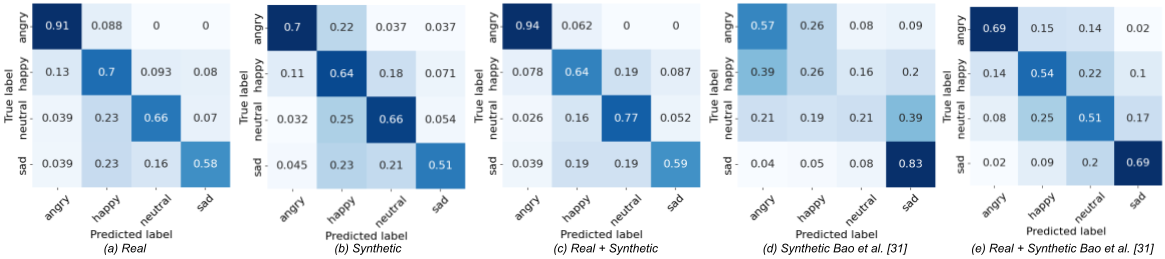}
\caption{Confusion Matrix Results for IEMOCAP data.}
\vspace{-1.5em}
\label{fig:cm}
\end{figure*}
In this experiment, we present SER results using the real, synthetic, and real+synthetic data in Table \ref{tab:within-corpus}. Results for IEMOCAP data are compared with recent studies \cite{sahu2018enhancing,bao2019cyclegan,latif2020augmenting}. In \cite{sahu2018enhancing}, the authors use conditional GANs to augment the ground truth training data with synthetic samples to improve speech emotion classification. In \cite{bao2019cyclegan}, the authors utilise a CycleGAN-based model for the augmentation of real data. In contrast to these studies, we achieve considerable improvement for synthetic and real+synthetic data. In \cite{latif2020augmenting}, the authors introduce a framework that utilises mixup augmentation while training a GAN-based network. They were able to improve SER performance by augmenting the training data. Our results are better compared to Latif et al., \cite{latif2020augmenting} without augmentation (see Table \ref{tab:within-corpus}). We achieve  further improvements in UAR ($61.38\pm2.04\,\%$) when mixup augmentation is applied to the training data similar to \cite{latif2020augmenting}. 

In Figure \ref{fig:cm}, we present the confusion matrices and compare the results with \cite{bao2019cyclegan} for synthetic and real+synthetic data cases. 
We 
find 
improved results for both cases. Most importantly, our results for synthetic data are consistent for all the classes compared to the \cite{bao2019cyclegan}, where they achieve high accuracy only on sad emotion and low on happy and neutral classes. This shows that our proposed model is capturing emotions and generating better emotional samples for all the classes.


\begin{table}[!ht]
\scriptsize
\centering
\caption{Results (UAR (\%)) on IEMOCAP for corpus setting.}
\begin{tabular}{|c|c|c|c|c|}
\hline
\textbf{Studies}  & \textbf{Real} & \textbf{Syn} & \textbf{Real+Syn} & \textbf{\tiny{Improvement}} \\ \hline
Sahu et al. \cite{sahu2018enhancing}       & 59.42         & 34.09         & 60.29               & 0.87                          \\ \hline
Bao et al. \cite{bao2019cyclegan}        & 59.48±0.71  & 46.59±0.75  & 60.37±0.70         & 0.89                          \\ \hline
Latif et al. \cite{latif2020augmenting}      & 60.51±0.57    & 45.75±0.81  & 61.05±0.68       & 0.54                          \\ \hline
Ours              &    58.62±2.11           & \textbf{57.96±1.54}            &  \textbf{61.22±1.85}                   &       \textbf{2.6}                        \\ \hline
\end{tabular}
\label{tab:within-corpus}
\end{table}

\subsubsection{Cross-corpus Evaluation}
In cross-corpus SER, we utilise the MSP-IMPROV corpus as target data. We perform experiments using real, synthetic, and real+synthetic data. To be consistent with studies \cite{bao2019cyclegan,latif2020augmenting,sahu2018enhancing} compared in this section, we randomly select 30\,\% of the samples from MSP-IMPROV as the development set for hyper-parameter tuning and the selection and the remaining 70\,\% as the test set.
\begin{table}[!ht]
\centering
\scriptsize
\caption{Comparing results for cross-corpus evaluation.}
\begin{tabular}{|l|l|l|l|}\hline
Studies      & Real         & Syn.         & Real+Syn.   \\\hline
Sahu et al. \cite{sahu2018enhancing}  & 45.14        & 33.96        & 45.40       \\\hline
Bao et al. \cite{bao2019cyclegan}   & 45.58 ± 0.40 & 41.58 ± 1.29 & 46.52±0.43  \\\hline
Latif et al. \cite{latif2020augmenting} & 46.0±0.57    & 42.15 ± 1.12 & 46.60 ±0.45 \\\hline
Our          & 45.81±0.65   & \textbf{43.53± 1.20}  & \textbf{48.22±0.51} \\\hline
Our (+mixup)          & \textbf{46.51±0.65}   & \textbf{44.31± 1.10}  & \textbf{48.58±0.51} \\\hline
\end{tabular}
\label{tab:cross-corpus}
\end{table}
Results are presented in Table \ref{tab:cross-corpus}. We compare the performance with different studies \cite{latif2020augmenting,sahu2018enhancing,bao2019cyclegan}. 
 All these studies \cite{sahu2018enhancing,bao2019cyclegan,latif2020augmenting} use GAN-based architectures to generate the synthetic features to augment the training of SER. We are achieving improved results compared to these studies for synthetic and real+synthetic data. However, we are achieving comparable results with \cite{latif2020augmenting} for real data. In \cite{latif2020augmenting}, the authors also utilise mixup augmentation to augment training data. We achieve better results with the utilisation of mixup augmentation in our approach (see Table \ref{tab:cross-corpus}). 

\begin{figure}[!ht]
\centering
\includegraphics[width=0.5\textwidth]{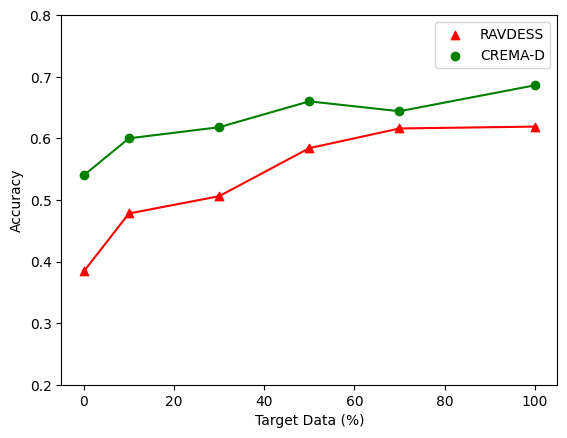}
\caption{Cross-Corpus results with varying percentages of target data in the training set.}
\vspace{-1.5em}
\label{fig:mix_plot}
\end{figure}

Most of the previous studies \cite{latif2020augmenting,bao2019cyclegan,sahu2018enhancing} performed cross-corpus evaluations on IEMOCAP and MSP-IMPROV. Both of these datasets are recorded in similar recording situations and have almost similar distributions. In this work, we extend our experiments to other datasets that have different distributions. We use CREMA-D and RAVDESS for these experiments. We trained our model on IEMOCAP synthetic data and evaluations are performed on 50\,\% of CREMA-D and RAVDESS. The remaining 50\,\% samples of these corpora are used as a presentation for model adaptation.
Results are presented in Figure \ref{fig:mix_plot}, which shows that synthetic data contains emotional information that helps the classifier to identify emotions across different datasets even when recorded in different situations. 



\section{Conclusions and Future Work}
In this work, we have addressed a major challenge of data scarcity in speech emotion recognition (SER) by proposing to use improved denoising diffusion probabilistic models (IDDPM) for synthetic data generation. We conditioned the IDDPM using the textual embedding from Bidirectional Encoder Representations to generate high-quality synthetic data. We used synthetic data to augment SER and the evaluations are performed both in within-corpus and cross-corpus settings using four publicly available datasets. In contrast to the recent studies on GAN-based synthetic data generation, our approach considerably helps improve SER performance with synthetic data utilisation for training data augmentation. In future works, we aim to design an extended version of the proposed framework for addressing the data scarcity issues in cross-lingual SER.




\end{document}